\documentclass[prb,twocolumn,superscriptaddress]{revtex4-2}
\usepackage[dvips]{graphicx}
\usepackage{type1cm}
\usepackage[usenames,svgnames]{xcolor}
\usepackage{color}
\usepackage{url}
\usepackage{natbib}
\usepackage[colorlinks=true,linkcolor=blue,citecolor=blue,breaklinks=true]{hyperref}

\usepackage{amsmath,amssymb,bm,amsthm,mathtools}

\DeclareMathOperator{\Tr}{\mathrm{Tr}}
\DeclareMathOperator{\re}{\mathrm{Re}}

\newcommand{\pd}{\partial}

\usepackage{braket}

\usepackage[capitalize]{cleveref}
\crefname{section}{Sec.}{Secs.}
\Crefname{section}{Section}{Sections}

\begin{document}
\nocite{apsrev41control}

\title{Liouvillian-gap analysis of open quantum many-body systems in the weak dissipation limit}

\author{Takashi Mori}
\affiliation{
RIKEN Center for Emergent Matter Science (CEMS), Wako 351-0198, Japan
}

\begin{abstract}
Recent experiments have reported that novel physics emerge in open quantum many-body systems due to an interplay of interactions and dissipation, which stimulate theoretical studies of the many-body Lindblad equation.
Although the strong dissipation regime receives considerable interest in this context, this work focuses on the weak bulk dissipation.
By examining the spectral property of the many-body Lindblad generator for specific models, we find that its spectral gap shows singularity in the weak dissipation limit when the thermodynamic limit is taken first.
Based on analytical arguments and numerical calculations, we conjecture that such a singularity is generic in bulk-dissipated quantum many-body systems and is related to the concept of the Ruelle-Pollicott resonance in chaos theory, which determines the timescale of thermalization of an isolated system.
This conjecture suggests that the many-body Lindblad equation in the weak dissipation regime contains nontrivial information on intrinsic properties of a quantum many-body system.
\end{abstract}
\maketitle

\section{Introduction}
Through recent experimental and theoretical studies, it is recognized that open quantum many-body systems exhibit intriguing nonequilibrium dynamics and steady states, which are different from those in isolated systems.
Because of the controllability of both the many-body Hamiltonian and dissipation, ultracold atoms provide an excellent experimental platform to study novel physics emerging in open quantum many-body systems, including entanglement generation~\citep{Barreiro2010,Krauter2011}, dissipative quantum phase transitions~\citep{Tomita2017}, anomalous decays of correlations~\citep{Bouganne2020}, continuous time crystals~\citep{Kongkhambut2022}, entanglement transitions under continuous quantum measurements~\citep{Noel2022,Koh2023}, to name a few.
Those interesting physics due to an interplay of many-body interactions and dissipation gives rise to the concept of dissipation engineering, which is an attempt of controlling quantum many-body systems by utilizing well-designed dissipation~\citep{Verstraete2009}.

Novel physics mentioned above typically emerge in the \emph{strong} dissipation regime.
In contrast, the \emph{weak} dissipation regime of open quantum many-body systems has been less explored, in spite of the fact that the Lindblad equation, which is a fundamental equation for Markovian open quantum systems, is usually justified in the weak dissipation regime~\citep{Breuer_text}.
The aim of this work is to draw attention to generic properties of many-body Lindblad equations in the weak dissipation regime.

Generally speaking, sufficiently weak dissipation enables us to probe intrinsic properties of a quantum many-body system.
For example, a recent work~\citep{Pan2020} proposed the non-Hermitian linear-response theory, which utilizes weak dissipation to get information about intrinsic correlations of a given quantum many-body system.
In this work, we show that the eigenvalue analysis of the Lindblad generator, which is often referred to as the Liouvillian or the Lindbladian in literature, in the weak dissipation limit leads to a better understanding of thermalization dynamics in \emph{isolated} quantum many-body systems.

One of the essential problems in the study of isolated quantum systems is to elucidate how irreversibility emerges from pure Hamiltonian dynamics.
Theoretical studies have clarified that concepts from chaos theory are helpful to describe thermalization of isolated quantum systems: the eigenstate thermalization hypothesis (ETH)~\citep{Deutsch1991,Srednicki1994,Rigol2008,DAlessio2016_review, Mori2018_review} is an important concept emerging from the quantum chaos.
The ETH states that every individual energy eigenstate is locally indistinguishable from thermal equilibrium, whose validity has been numerically tested for various nonintegrable models~\citep{Kim2014,Beugeling2014,Beugeling2015}.

Although the ETH provides sufficient criteria of quantum ergodicity, it does not tell us much about the timescale of the onset of thermalization.
Since the time evolution operator is unitary, all of its eigenvalues lie on the unit circle in the complex plane.
When the system size increases, the eigenvalue spectrum becomes more and more dense, and such quasi-continuous spectrum is responsible for irreversible relaxation.
It is however difficult to gain insights on the thermalization timescale from the quasi-continuous spectrum on the unit circle.
So far, there are some previous attempts to figure out the thermalization timescale along the typicality approach~\citep{Goldstein2015_extremely, Reimann2016}, where we consider a random Hamiltonian, instead of analyzing a concrete system.
However, the typicality approach is too general and often fails to give a correct estimate of the thermalization time, especially in systems with slow relaxation called prethermalization~\citep{Mori2018_review}.

The main message of this work is that eigenvalue analysis of the dissipative Liouvillian in the weak dissipation limit allows us to extract exponential decays from the unitary dynamics of the isolated system.
More precisely, by examining the operator-spreading dynamics, we argue that the spectral gap of the Liouvillian (or its variant) does not simply vanish even in the weak dissipation limit \emph{when the thermodynamic limit is taken first}.
We conjecture that this nonzero Liouvillian gap in the weak dissipation limit is related to the timescale of thermalization of the isolated system.
We also discuss the relation to the Ruelle-Pollicott (RP) resonance~\citep{Ruelle1986, Pollicott1985}, which is a fundamental concept in the theory of classical chaos~\citep{Gaspard_text,Dorfman_text}.
The conjecture is verified by numerical calculations.

The Liouvillian gap has been studied intensively so far because it is an essential quantity characterizing an open quantum system~\citep{Kessler2012,Cai2013,Kastoryano2013,Znidaric2015,Casteels2017,Minganti2018,Shibata2019,Shibata2019_dissipative,Mori2023_symmetrized,Shirai2023_accelerated}.
This work sheds new light on this problem by revealing a general connection between the many-body Liouvillian gap and the chaotic property of an underlying isolated quantum system.

The rest of this work is organized as follows.
In \cref{sec:setup}, we describe theoretical setup of open quantum many-body systems.
In \cref{sec:Floquet}, we discuss the generic feature of the Liouvillian eigenvalue spectrum in the weak-dissipation limit.
Based on an analogy with the RP resonance in classical chaos, we present our conjecture that the Liouvillian gap in the weak-dissipation limit, which should be taken after the thermodynamic limit, gives the intrinsic decay rate of the isolated system.
The conjecture is verified by numerical calculations for the kicked Ising model under bulk dephasing.
In \cref{sec:static}, we discuss the Liouvillian eigenvalue spectrum in static systems.
In contrast to Floquet systems, the Liouvillian gap in a static system is not related to the intrinsic decay rate of the isolated system, but it turns out that some other eigenvalues of the Liouvillian can be interpreted as quantum RP resonances.
We introduce a projection super-operator, which correctly picks up Liouvillian eigenvalues that can be interpreted as quantum RP resonances, and argue that the spectral gap of the \emph{projected Liouvillian} gives the intrinsic decay rate of the isolated static system.
This is also numerically verified in a specific model.
In \cref{sec:conclusion}, we conclude this work with with a summary and an outlook.

\section{Theoretical setup}
\label{sec:setup}

\begin{figure}
\centering
\includegraphics[width=0.9\linewidth]{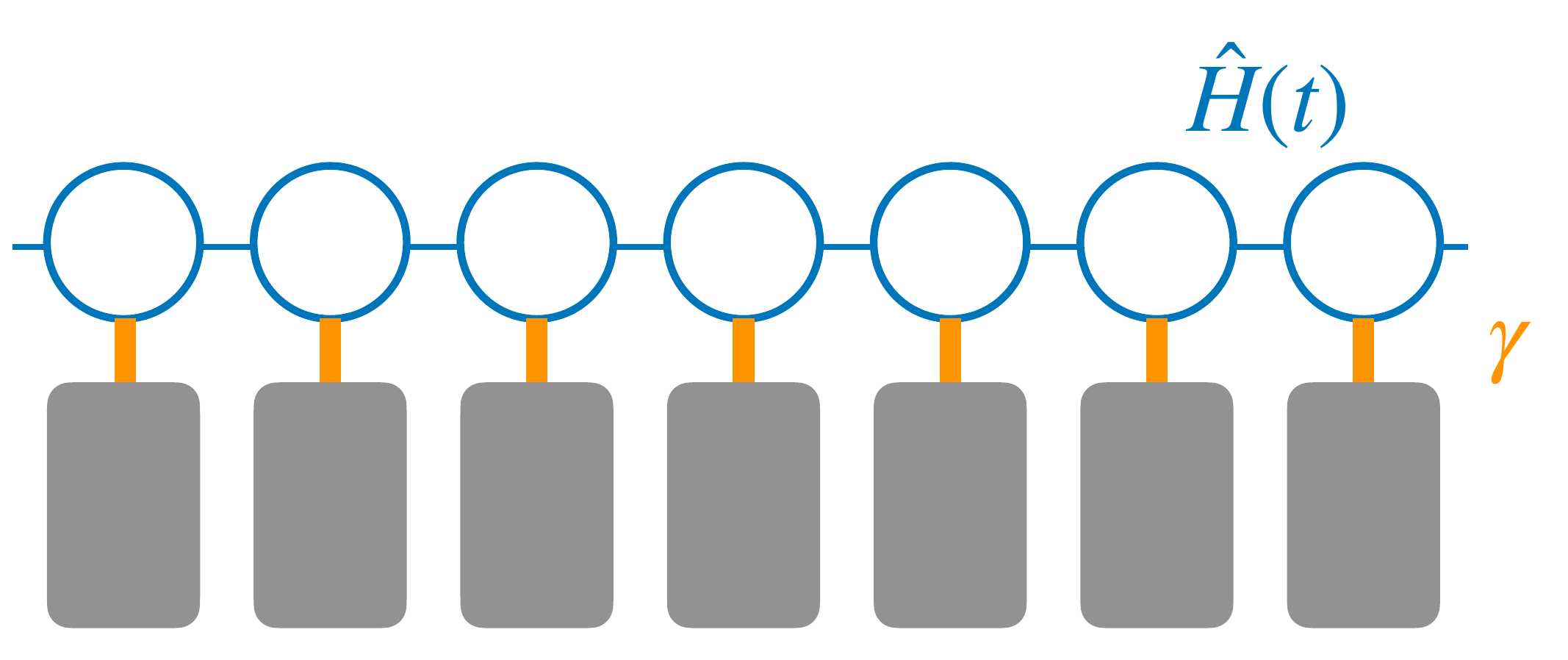}
\caption{Theoretical setting. We consider an open quantum many-body system, which is described by $\hat{H}(t)$, under bulk dissipation of strength $\gamma$.}
\label{fig:setting}
\end{figure}

In this work, we consider quantum many-body systems under bulk dissipation.
To be specific, we consider a spin-1/2 chain with $L$ sites, although our argument is not restricted to this specific model.
Each site is labeled by $i=1,2,\dots,L$, and Pauli operators at site $i$ are denoted by $\hat{\sigma}_i^\alpha$ ($\alpha=x,y,z$).

Let us denote by $\hat{H}(t)$ the Hamiltonian of the system that may explicitly depend on time with period $\tau$: $\hat{H}(t)=\hat{H}(t+\tau)$.
The static system is included as a special case.
The Markovian bulk dissipation is described by the following quantum master equation of the Lindblad form~\citep{Mori2023_review}:
\begin{align}
\frac{d\rho}{dt}&=-i[\hat{H}(t),\rho]+\gamma\sum_{i=1}^L\left(\hat{L}_i\rho\hat{L}_i^\dagger-\frac{1}{2}\{\hat{L}_i^\dagger\hat{L}_i,\rho\}\right)
\nonumber \\
&\eqqcolon\mathcal{L}(t)\rho,
\label{eq:Lindblad}
\end{align}
where $\rho(t)$ is the density matrix at time $t$, $\gamma>0$ stands for the strength of dissipation, and $\hat{L}_i$ is a jump operator at site $i$, which characterizes the type of dissipation.
\Cref{fig:setting} illustrates the theoretical setting.

In this work, we mainly consider bulk dephasing corresponding to $\hat{L}_i=\hat{\sigma}_i^z$, but our main result does not depend on this specific choice of jump operators, which is demonstrated in \cref{sec:other}.

The superoperator $\mathcal{L}(t)$ defined in \cref{eq:Lindblad} is called the Liouvillian or the Lindbladian.
For static systems with no time dependence, $\mathcal{L}(t)=\mathcal{L}$, we denote by $\{\lambda_\alpha\}$ eigenvalues of $\mathcal{L}$.
The Liouvillian has zero eigenvalue $\lambda_0=0$ that corresponds to the steady state.
For simplicity, we assume that the zero eigenvalue is not degenerate: the system has a unique steady state.
It is shown that the real part of any eigenvalue is nonpositive $\re\lambda_\alpha\leq 0$.
The Liouvillian gap $g$ is defined as the largest real part of nonzero eigenvalues:
\begin{align}
g=-\max_{\alpha\neq 0}\re\lambda_\alpha.
\label{eq:gap}
\end{align}

In open Floquet systems with the time-periodic Liouvillian $\mathcal{L}(t)=\mathcal{L}(t+\tau)$, we define the Liouvillian gap in the following way.
We introduce the time evolution operator over one cycle (the Floquet operator)
\begin{align}
\mathcal{U}_\mathrm{F}=\mathcal{T}e^{\int_0^\tau\mathcal{L}(t)dt},
\end{align}
where $\mathcal{T}$ denotes the time-ordering operation.
Let us denote by $e^{\lambda_\alpha\tau}$ the eigenvalues of $\mathcal{U}_\mathrm{F}$.
We simply call $\{\lambda_\alpha\}$ ``Liouvillian eigenvalues'' in Floquet systems.
The unique periodic steady state corresponds to the right eigenmode with zero eigenvalue $\lambda_0=0$.
Any other eigenvalue has a non-positive real part, where $-\re\lambda_\alpha$ is the decay rate of the corresponding eigenmode.
The Liouvillian gap is then defined by \cref{eq:gap}, which is the smallest decay rate among all the non-stationary eigenmodes.

In previous works, the Liouvillian gap has been investigated for various models because it has some important properties.
Firstly, the Liouvillian gap gives the \emph{asymptotic} decay rate of the open system~\citep{Kessler2012}.
It is obvious that the slowest eigenmode is dominant in the long-time limit.
Strictly speaking, the Liouvillian gap gives a lower bound on the asymptotic decay rate.
Here, it is remarked that the Liouvillian gap in general does not give a lower bound on the decay rate in a \emph{transient} regime~\citep{Znidaric2015,Mori2020_resolving,Haga2021}.
The symmetrized Liouvillian gap, which is a variant of the usual Liouvillian gap, gives a correct lower bound~\citep{Mori2023_symmetrized,Shirai2023_accelerated}.
Secondly, the Liouvillian gap is related to the property of the steady state.
It was shown that a finite Liouvillian gap in the thermodynamic limit implies exponential decays of spatial correlations in the steady state~\citep{Kastoryano2013}.
Thirdly, the closing of the Liouvillian gap in the thermodynamic limit is a signature of dissipative phase transitions~\citep{Casteels2017,Minganti2018}.
In this way, the Liouvillian gap is a fundamental quantity, which contains rich physical information.

For later convenience, it is useful to introduce the Liouvillian in the Heisenberg picture.
For any operator $\hat{A}$, its expectation value at time $t$ is written as $\braket{\hat{A}(t)}=\Tr[\hat{A}\rho(t)]$, where $\rho(t)$ obeys \cref{eq:Lindblad}.
The same quantity is also expressed as $\braket{\hat{A}(t)}=\Tr[\hat{A}(t)\rho(0)]$, where $\hat{A}(t)$ obeys 
\begin{align}
\frac{d\hat{A}}{dt}&=i[\hat{H}(t),\hat{A}(t)]
+\gamma\sum_{i=1}^L\left(\hat{L}_i^\dagger\hat{A}(t)\hat{L}_i-\frac{1}{2}\{\hat{L}_i^\dagger\hat{L}_i,\hat{A}(t)\}\right)
\nonumber \\
&\eqqcolon\mathcal{L}^\dagger(t)\hat{A}(t).
\label{eq:Heisenberg}
\end{align}
Here, $\mathcal{L}^\dagger(t)$ is the Liouvillian in the Heisenberg picture.
This notation stems from the fact that $\mathcal{L}^\dagger(t)$ is the Hermitian conjugate of $\mathcal{L}(t)$ under the inner product $\braket{\hat{A},\hat{B}}=\Tr[\hat{A}^\dagger\hat{B}]$, i.e., $\braket{\hat{A},\mathcal{L}(t)\hat{B}}=\braket{\mathcal{L}^\dagger(t)\hat{A},\hat{B}}$.

Correspondingly, we define the Floquet operator in the Heisenberg picture $\mathcal{U}_\mathrm{F}^\dagger=\bar{\mathcal{T}}e^{\int_0^\tau\mathcal{L}^\dagger(t)dt}$ ($\bar{\mathcal{T}}$ is the anti-time ordering).
By using the Hermiticity preserving property of the Liouvillian, $(\mathcal{L}(t)\hat{A})^\dagger=\mathcal{L}(t)\hat{A}^\dagger$, it is shown that the eigenvalue spectrum of $\mathcal{U}_\mathrm{F}^\dagger$ coincides with that of $\mathcal{U}_\mathrm{F}$.

In the following sections, we investigate the weak dissipation limit $\gamma\to +0$.
We emphasize that the limit of $\gamma\to +0$ should be taken after taking the thermodynamic limit.
If we take the limit of $\gamma \to +0$ first, the Liouvillian becomes anti-Hermitian, $\lim_{\gamma\to +0}\mathcal{L}(t)=-i[\hat{H}(t),\cdot]$, and the Liouvillian gap is trivially zero.
We will see that the interplay of the thermodynamic limit and the limit of weak bulk dissipation leads to unexpected behavior.

\section{Main idea for Floquet systems}
\label{sec:Floquet}
In this section, the main result is presented for open many-body Floquet systems.
Since a special care is needed for static systems, discussion on static systems is postponed until \cref{sec:static}.

The following argument is valid for generic lattice systems, but for concreteness, we consider the kicked Ising chain under bulk dephasing.
The Hamiltonian is given by $\hat{H}(t)=\hat{H}_0+\hat{V}(t)$, where
\begin{align}
\left\{
\begin{aligned}
&\hat{H}_0=-\sum_{i=1}^L\left(J\hat{\sigma}_i^z\hat{\sigma}_{i+1}^z+h_z\hat{\sigma}_i^z\right);\\
&\hat{V}(t)=-h_x\tau\sum_{n=-\infty}^\infty\delta(t-n\tau)\sum_{i=1}^L\hat{\sigma}_i^x.
\end{aligned}
\right.
\label{eq:H_kick}
\end{align}
The bulk dephasing is expressed by jump operators
\begin{align}
\hat{L}_i=\hat{\sigma}_i^z
\end{align}
for each site $i=1,2,\dots,L$.
In numerical calculations, we fix $J=1$, $h_z=0.8090$, $h_x=0.9045$ throughout this section, and consider three different values of $\tau$: $\tau=0.65$, $0.7$, and $0.75$.

\subsection{Operator growth under unitary time evolution}
\label{sec:operator_growth}
We begin our discussion with the unitary time evolution of an Hermitian operator $\hat{A}(t)$ in the Heisenberg picture, which obeys \cref{eq:Heisenberg} with $\gamma=0$, i.e., $d\hat{A}(t)/dt=i[\hat{H}(t),\hat{A}(t)]$.
The effect of dissipation is taken into account in \cref{sec:singularity}.

Due to spin-spin interactions, even if $\hat{A}(0)$ is a local operator, $\hat{A}(t)$ at $t>0$ will spread over the system.
In order to describe such operator growth, we introduce the notion of the average operator size $S[\hat{A}]$ of $\hat{A}$~\citep{Nahum2018, Roberts2018, Yin2020, Schuster2022}.
It is given by
\begin{align}
S[\hat{A}]\coloneqq \frac{\Tr[\hat{A}\mathcal{S}(\hat{A})]}{\Tr[\hat{A}^2]},
\label{eq:operator_size}
\end{align}
where
\begin{align}
\mathcal{S}(\hat{A})=\frac{1}{4}\sum_{i=1}^L\sum_{\alpha=x,y,z\}}(\hat{A}-\hat{\sigma}_i^\alpha\hat{A}\hat{\sigma}_i^\alpha).
\end{align}
According to this definition, a Pauli string $\hat{\sigma}_{i_1}^{\alpha_1}\hat{\sigma}_{i_2}^{\alpha_2}\dots\hat{\sigma}_{i_\ell}^{\alpha_\ell}$ with $1\leq i_1<i_2<\dots<i_\ell<L$ and $\alpha_i\in\{x,y,z\}$ has the operator size $\ell$.
The average operator size of $\sigma_i^x\sigma_j^x+\sigma_k^z$ with $i\neq j$, for example, is $1.5$.
The relation between the average operator size of \cref{eq:operator_size} and out-of-time-ordered correlations (OTOCs)~\citep{Larkin1969, Shenker2014, Maldacena2016} is discussed in Refs.~\citep{Nahum2018, Roberts2018, Schuster2022}.

In a short-range interacting spin chain, $S[\hat{A}(t)]$ usually increases in a linear way: $S[\hat{A}(t)]\sim vt$ with $v>0$ being a constant that is independent of the system size (its upper bound is given by the Lieb-Robinson velocity~\citep{Lieb1972}).
Eventually, the operator will spread over the entire system for $t\gtrsim L/v$, and the operator size will be saturated at a value that is proportional to $L$, which is called the operator scrambling.

\subsection{Singularity at $\gamma=0$}
\label{sec:singularity}
In the absence of dissipation, the time evolution is unitary and Liouvillian eigenvalues $\{\lambda_\alpha\}$ are pure imaginary.
Therefore, $g=0$ at $\gamma=0$ for any finite system.
It would be a bit surprise if we have $g>0$ in the limit of $\gamma\to +0$, which is indeed the case as we argue below.

In the previous subsection, we discuss the operator spreading under the unitary time evolution.
Now we consider the effect of very weak bulk dissipation $\gamma>0$ to this operator dynamics.
In this section, we assume that the system is intrinsically chaotic and has no local conserved quantity when it is completely isolated from the environment (i.e. $\gamma=0$).
When $\gamma t\ll 1$, it is expected that dissipation has essentially no effect.
Therefore, if $\gamma\ll v/L$, the system undergoes dissipative relaxation \emph{after the saturation of the operator size} $S[\hat{A}(t)]\sim L$.
A crucial observation is that the influence of the bulk dissipation on $\hat{A}(t)$ is proportional to its operator size $S[\hat{A}(t)]$~\citep{Shirai2023_accelerated}.
It is understood from the fact that the dissipator $\mathcal{D}=\sum_{i=1}^L\mathcal{D}_i$ of \cref{eq:Heisenberg} with
\begin{align}
\mathcal{D}_i\hat{A}=\gamma\left(\hat{L}_i^\dagger\hat{A}\hat{L}_i-\frac{1}{2}\{\hat{L}_i^\dagger\hat{L}_i,\hat{A}\}\right)
\end{align}
has the property $\mathcal{D}_i\hat{A}=0$ if $[\hat{A},\hat{L}_i]=[\hat{A},\hat{L}_i^\dagger]=0$.
If the support of $\hat{A}$ does not contain the site $i$, $\mathcal{D}_i\hat{A}$ vanishes.
As a result, we have
\begin{align}
\frac{\|\hat{D}\hat{A}\|}{\|\hat{A}\|}=\frac{\left\|\sum_{i=1}^L\mathcal{D}_i\hat{A}\right\|}{\|\hat{A}\|}\sim \gamma S[\hat{A}].
\end{align}
Since $S[\hat{A}(t)]\sim L$ after the operator scrambling, the effective dissipation strength for $\hat{A}(t)$ is proportional to $\gamma L$.
This conclusion is valid for an arbitrary local operator $\hat{A}$ under the assumption that the many-body Hamiltonian $\hat{H}(t)$ is chaotic and has no local conserved quantity.
It implies that the asymptotic decay rate, which is nothing but the Liouvillian gap $g$, is also proportional to $\gamma L$:
\begin{align}
g\propto \gamma L \quad\text{for }\gamma\ll\frac{v}{L}.
\end{align}

When $v/L\ll\gamma\ll v$, dissipation takes place before the operator spreads over the entire system.
The operator size approximately grows without dissipation up to $t_\gamma\sim\gamma^{-1}$ and reaches $S[\hat{A}(t_\gamma)]\sim vt_\gamma\sim\gamma^{-1}$.
Afterwards, the operator-size growth stops due to dissipation: $S[\hat{A}(t)]\approx S[\hat{A}(t_\gamma)]$ for $t>t_\gamma$.
As a result, the asymptotic decay rate is given by
\begin{align}
g\sim\gamma S[\hat{A}(t_\gamma)]\sim 1,
\end{align}
which is independent of $L$ and $\gamma$.

The above argument implies discontinuity of the Liouvillian gap at $\gamma=0$ in the thermodynamic limit.
We have argued that there are two different regimes: $g\sim\gamma L$ for $\gamma\lesssim v/L$ and $g\sim 1$ for $v/L\ll\gamma\ll v$.
In the thermodynamic limit, the former regime disappears.
We therefore expect that
\begin{align}
\lim_{\gamma\to +0}\lim_{L\to\infty}g\eqqcolon \bar{g}>0.
\label{eq:g_bar}
\end{align}
The Liouvillian gap remains finite in the weak dissipation limit if the thermodynamic limit is taken first.
As we have already mentioned, we have $g=0$ if we put $\gamma=0$ before the thermodynamic limit.
\Cref{eq:g_bar} thus manifests a singularity (discontinuity) of the Liouvillian gap in the thermodynamic limit, which is a generic feature of chaotic open many-body Floquet systems.

It should be noted that a similar observation was reported in the dissipative Sachdev-Ye-Kitaev model~\citep{Sa2022,Garcia-Garcia2023}.

\begin{figure}[t]
\centering
\includegraphics[width=0.9\linewidth]{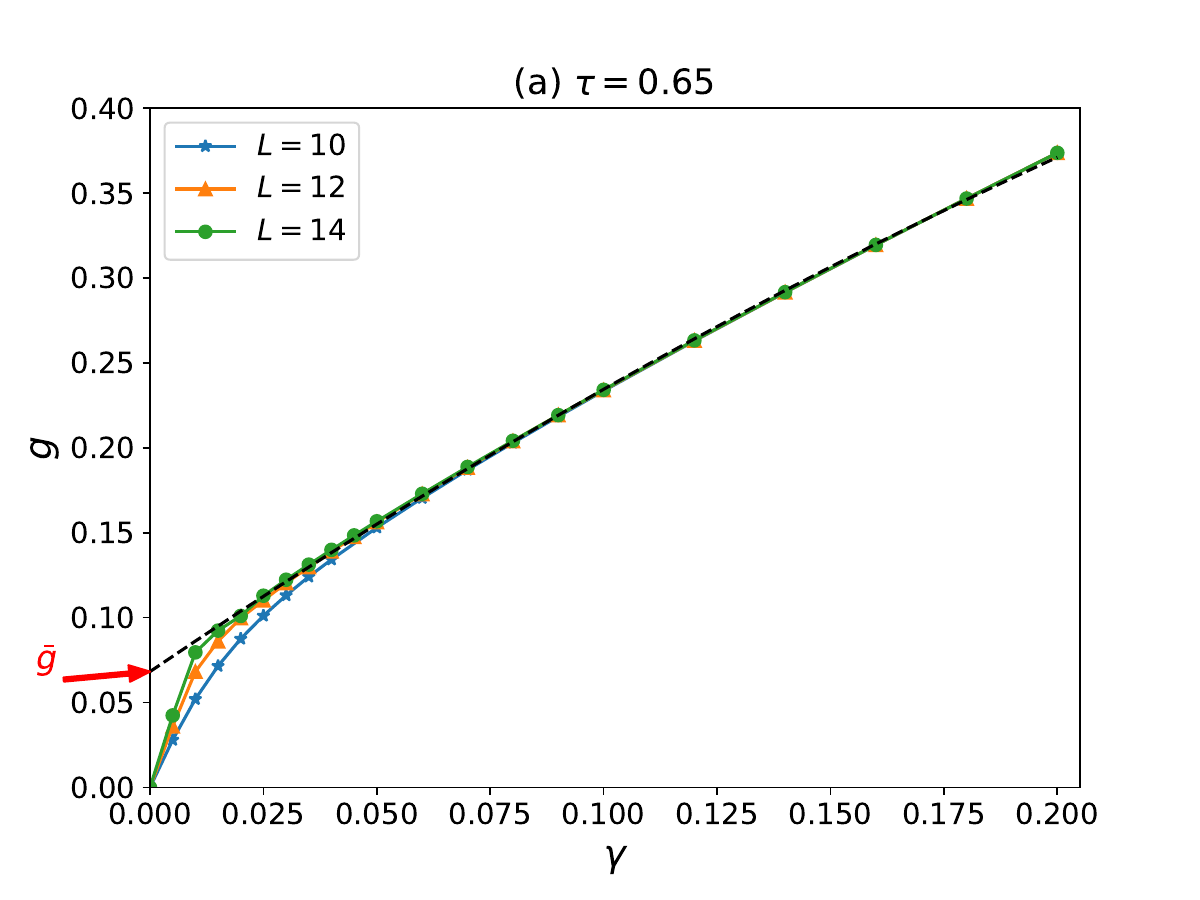}
\includegraphics[width=0.9\linewidth]{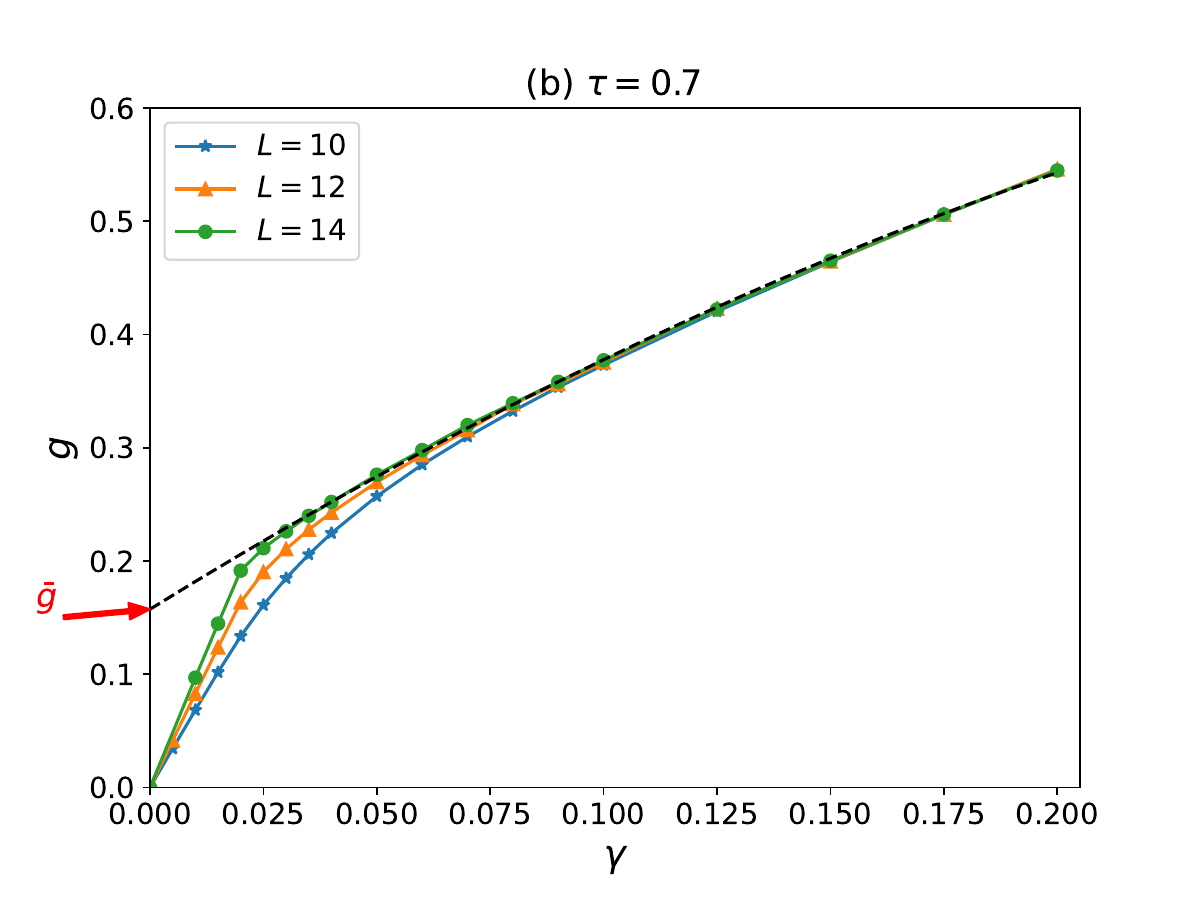}
\includegraphics[width=0.9\linewidth]{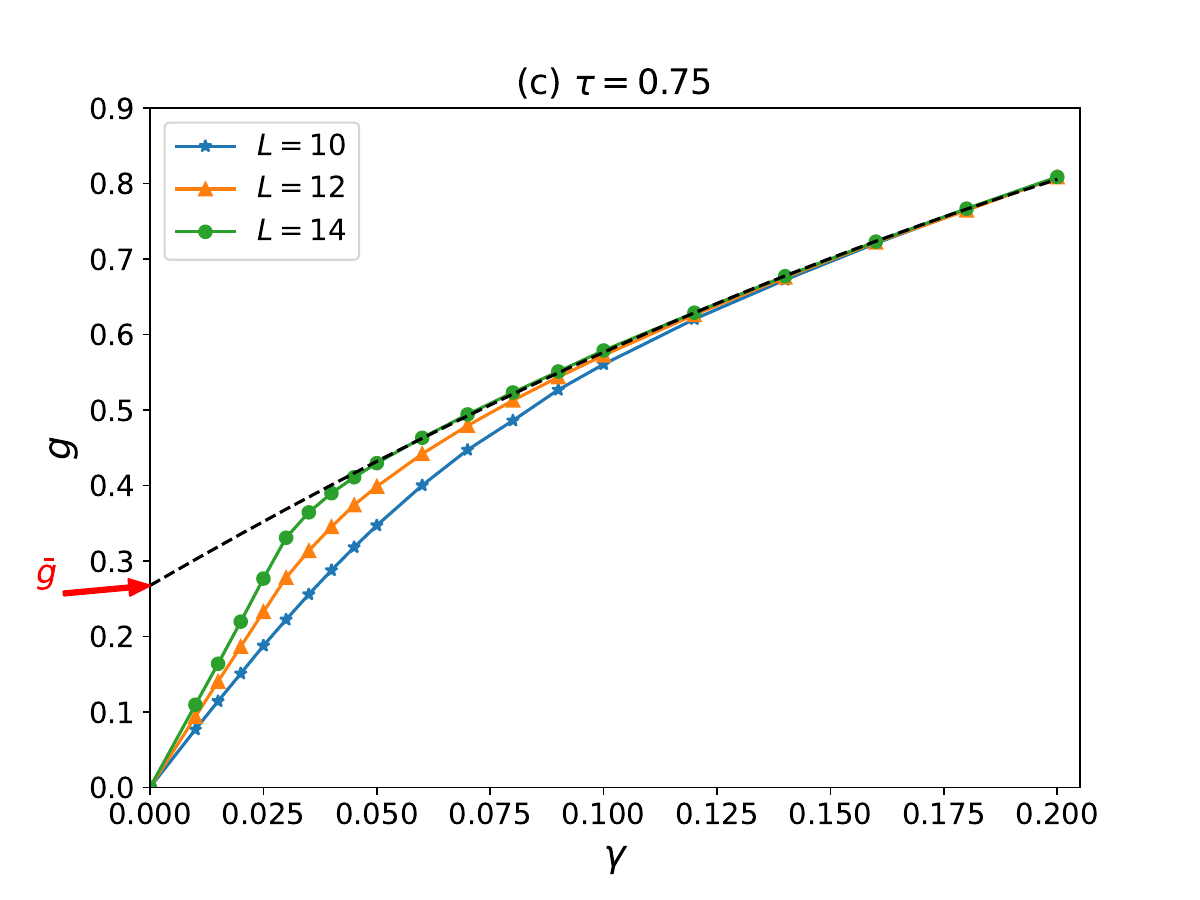}
\caption{Liouvillian gap as a function of $\gamma$ for $L=10$, $12$, and $14$.
We consider three different values of $\tau$: (a) $\tau=0.65$, (b) $\tau=0.7$, and (c) $\tau=0.75$.
Dashed lines are quadratic functions fitted to the data of $L=14$ with $\gamma$ greater than a certain value (see the text).}
\label{fig:singularity}
\end{figure}

This theoretical prediction is validated by numerical calculations for the kicked Ising chain under bulk dephasing.
Numerical results on the Liouvillian gap are presented in \cref{fig:singularity} for $\tau=0.65$, $0.7$, and $0.75$.
We see that the Liouvillian gap increases with $L$ at a sufficiently small $\gamma$, but is almost independent of $L$ for larger $\gamma$, which is consistent with the above theoretical prediction.
The dashed lines in \cref{fig:singularity} are quadratic functions fitted to the data of (a) $\gamma\geq 0.015$, (b) $\gamma\geq 0.03$, and (c) $\gamma\geq 0.05$ for $L=14$.
Extrapolated values to $\gamma=0$ are our numerical estimates of $\bar{g}$.
We obtain $\bar{g}=0.0682$, $0.157$, and $0.268$ for $\tau=0.65$, $0.7$, and $0.75$, respectively.

The reason why a quadratic fitting is used is that an accurate extrapolation requires numerical data for relatively large values of $\gamma$ ($\lesssim 0.2$) in the largest system size $(L=14)$ accessible in our numerical calculations.
If we could perform numerical calculations for larger system sizes, an accurate extrapolation using numerical data for smaller values of $\gamma$ would be possible, and then a linear fitting would be enough.

\subsection{Ruelle-Pollicott resonances}
\label{sec:RP_resonance}
In \cref{sec:singularity}, we have seen that the Liouvillian gap shows singularity at $\gamma=0$: the Liouvillian gap converges to a nonzero value $\bar{g}$ in the weak dissipation limit.
A natural question is what its physical meaning is.
In \cref{sec:connection}, we will argue that  $\bar{g}$ is interpreted as a quantum RP resonance.
Before that, we begin with a brief exposition of RP resonances of the Hamiltonian dynamics.

Suppose a classical system with the Hamiltonian $H_t(\Gamma)$, where 
\begin{align}
\Gamma=(q_1,q_2,\dots,q_N,p_1,p_2,\dots,p_N)
\end{align}
denotes a point in the classical phase space.
Here, $\{q_i\}$ and $\{p_i\}$ are canonical variables, and the Hamiltonian may explicitly depend on $t$ with period $\tau$, i.e., $H_t(\Gamma)=H_{t+\tau}(\Gamma)$.
A dynamical trajectory is given by a one-parameter family $\Gamma_t$ of phase-space points, where the canonical variables obey the Hamilton equations of motion: $\dot{q}_i=\pd H(\Gamma_t)/\pd p_i$ and $\dot{p}_i=-\pd H(\Gamma_t)/\pd q_i$.
Let us denote by $P(\Gamma)$ the phase-space probability density.
Its time evolution $P_t(\Gamma)$ is formally expressed as
\begin{align}
P_t(\Gamma)=P(\Gamma_{-t})\eqqcolon \mathcal{U}_tP(\Gamma),
\end{align}
where $\Gamma=\Gamma_0$ and $\mathcal{U}_t$ is the time evolution operator of the phase-space density, which is referred to as the Frobenius-Perron operator~\citep{Gaspard_text,Dorfman_text}.
We can generally write $\mathcal{U}_t=\mathcal{T}e^{\int_0^t\mathcal{L}_\mathrm{cl}(t')dt'}$, where the generator $\mathcal{L}_\mathrm{cl}(t)$ is the classical Liouvillian in the absence of dissipation:
\begin{align}
\mathcal{L}_\mathrm{cl}(t)=\{H_t(\Gamma),\cdot\}_\mathrm{PB},
\end{align}
where $\{\cdot,\cdot\}_\mathrm{PB}$ denotes the Poisson bracket.
It corresponds to $-i[\hat{H}(t),\cdot]$ in quantum mechanics.

The classical Liouvillian is anti-Hermitian, i.e., $\braket{f,\mathcal{L}_\mathrm{cl}(t)g}=-\braket{\mathcal{L}_\mathrm{cl}(t)f,g}$, under the inner product
\begin{align}
\braket{f,g}\coloneqq\int d\Gamma\, f(\Gamma)^*g(\Gamma),
\end{align}
where $d\Gamma=dq_1dq_2\dots dq_Ndp_1dp_2\dots dp_N$.
It means that the Frobenius-Perron operator $\mathcal{U}_t$ is unitary within the Hilbert space of square-integrable functions, which expresses the reversibility of the Hamiltonian dynamics.

Because of the unitarity, the Floquet operator $\mathcal{U}_\tau$ has a spectrum on the unit circle in the complex plane.
Moreover, $\mathcal{U}_t$ has a continuous spectrum when the dynamics is chaotic.
With this in mind, let us consider the resolvent of the Floquet operator
\begin{align}
R(z)=(z-\mathcal{U}_\tau)^{-1}.
\end{align}
The resolvent is regularly behaved for any $z$ with $|z|\neq 1$.
The continuous spectrum of $\mathcal{U}_t$ gives a branch cut of $R(z)$, and hence we can perform an analytic continuation from $|z|>1$ to $|z|<1$ through the continuous spectrum~\citep{Gaspard_text,Hasegawa1992}.
It turns out that $R(z)$ may have poles inside the unit circle in the second Riemann sheet.
Those poles are written as $e^{\nu_i\tau}$ with $\re\nu_i<0$, and $\{\nu_i\}$ are known as RP resonances~\citep{Ruelle1986,Pollicott1985,Hasegawa1992,Gaspard_text}.
RP resonances are not true eigenvalues in the Hilbert space of square-integrable functions, but are understood as generalized eigenvalues in a wider Hilbert space~\citep{Gaspard_text}.

By using RP resonances, we can decompose the time evolution of the expectation value $\braket{A}_t=\int d\Gamma\, A(\Gamma)P_t(\Gamma)$ of a physical quantity $A(\Gamma)$ into the sum of exponential decays as follows:
\begin{align}
\braket{A}_t\sim \sum_{\nu_i\in\text{RP resonances}}C_ie^{\nu_i t},
\end{align}
where $t=n\tau$ with $n$ being an integer and we assume $\lim_{t\to\infty}\braket{A}_t=0$.
The long-time limit is governed by the leading RP resonance $\nu_*$ (the RP resonance with the largest real part):
\begin{align}
\braket{A}_t\sim e^{\nu_*t} \quad (t\to\infty).
\end{align}
RP resonances thus extract exponential decays hidden in the unitary time evolution.

Numerically, RP resonances can be obtained by diagonalizing a coarse-grained Frobenius-Perron operator~\citep{Weber2000,Klus2016_review}.
In numerical calculations, we discretize the phase space and express the Frobenius-Perron operator as a finite-dimensional matrix.
This discretization procedure corresponds to a coarse graining, which makes the Frobenius-Perron operator non-unitary.
Therefore, if we write eigenvalues of the Frobenius-Perron operator as $e^{\lambda_i\tau}$, where we call $\{\lambda_i\}$ (classical) Liouvillian eigenvalues, $\lambda_i$ have negative real part.
Interestingly, some of them still have negative real part even in the limit of the continuous phase space~\citep{Weber2000}.
It means that even infinitesimal coarse graining drastically changes the Liouvillian eigenvalues.
Those Liouvillian eigenvalues with negative real part in the continuous limit are nothing but RP resonances.

A natural question is whether quantum analogs of RP resonances exist.
As we have discussed above, a continuous spectrum of the Liouvillian for the pure Hamiltonian dynamics is crucial to get RP resonances (recall that the continuous spectrum plays the role of a branch cut of the resolvent, and an analytic continuation of the resolvent through the continuous spectrum brings about RP resonances as poles in the second Riemann sheet).
In classical systems, chaos ensures the existence of a continuous spectrum, whereas any finite quantum system has only discrete energy eigenvalues.
In quantum mechanics, a continuous spectrum appears in the semiclassical limit, and hence quantum analogs of RP resonances may exist in this limit.
Indeed, some previous works found RP resonances by considering the semiclassical limit of quantum systems~\citep{Pance2000,Garcia-Mata2003,Manderfeld2003}.

There is another limit that leads to a continuous spectrum in quantum mechanics: that is the \emph{thermodynamic limit}.
It is much less trivial whether there exist quantum RP resonances in a quantum many-body system that is far from any classical limit.
This problem has rarely been investigated in the literature, except for Refs.~\citep{Prosen2002,Prosen2004}, where Prosen found quantum RP resonances of the kicked Ising chain by introducing an appropriate coarse graining procedure in the operator space.

It should be noted that RP resonances differ from Lyapunov exponents: the former is concerned with long-time behavior, whereas the latter with short-time behavior.
Indeed, \citet{Garcia-Mata2018} showed that the long-time decay of OTOCs in a semiclassical model gives a quantum RP resonance, whereas Lyapunov exponents are related with short-time behavior of OTOCs.

\subsection{Connection between Liouvillian gap and Ruelle-Pollicott resonances}
\label{sec:connection}

\begin{figure}
\centering
\includegraphics[width=0.95\linewidth]{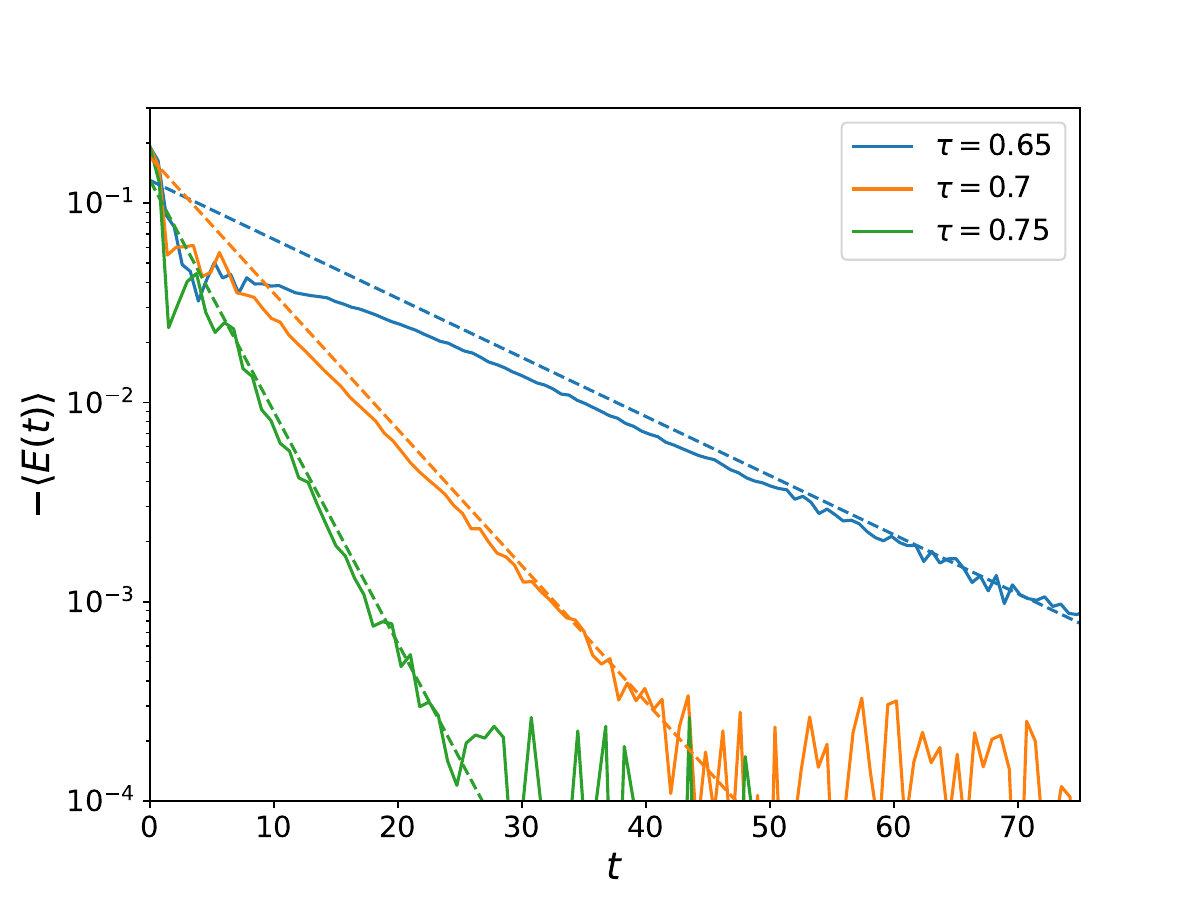}
\caption{Comparison between numerical solutions of the Schr\"odinger equation in the kicked Ising model with $L=28$ (solid lines) and exponential decays $\propto e^{-\bar{g}t}$ predicted by the Liouvillian gap analysis in \cref{fig:singularity} (dashed lines).}
\label{fig:exp_decay_Floquet}
\end{figure}

As we have seen in \cref{sec:RP_resonance}, RP resonances are related to instabilities against infinitesimal coarse graining.
Instead of performing a coarse graining, we can mimic it by introducing stochastic noise to the deterministic Hamiltonian dynamics.
As we explain below, RP resonances are also regarded as a manifestation of instabilities against infinitesimally weak stochastic noise~\citep{Gaspard1995,Khodas2000,Gaspard2002} (also see Ref.~\citep{Kurchan2009_lecture} for a pedagogical exposition on this and related topic).

When white Gaussian noise is added to the Hamiltonian dynamics, the dynamics is described by the Langevin equation.
The corresponding time evolution equation of the probability density $P_t(\Gamma)$ is given by the Kramers equation (or the Fokker-Planck equation in the overdamped limit): $\pd P_t(\Gamma)/\pd t=\mathcal{L}_K(t)P_t(\Gamma)$ with $\mathcal{L}_K(t)=\mathcal{L}_K(t+\tau)$.
The eigenvalues of the corresponding Floquet operator $\mathcal{U}_K=\mathcal{T}e^{\int_0^\tau\mathcal{L}_K(t)dt}$ are expressed as $e^{\lambda_i\tau}$.
Here, $\{\lambda_i\}$ are Liouvillian eigenvalues in the classical dynamics under stochastic noise.
In the presence of noise, Liouvillian eigenvalues have negative real part.
Surprisingly, some Liouvillian eigenvalues still have negative real part even in the weak noise limit.
Those nontrivial Liouvillian eigenvalues in the weak noise limit are nothing but RP resonances of the deterministic Hamiltonian dynamics.

Now an analogy between this situation and our Liouvillian gap analysis in open quantum systems would be obvious.
What we have found in \cref{sec:singularity} is that if we add weak dissipation to the Schr\"odinger dynamics of a quantum many-body system, Liouvillian eigenvalues have negative real part even in the weak dissipation limit.
Considering the analogy with classical chaotic systems, it is natural to conjecture that \emph{$\bar{g}$ in \cref{eq:g_bar} corresponds to (the real part of) the leading RP resonance of the isolated quantum many-body system}.

A numerical evidence of this interpretation for the kicked Ising model is presented in \cref{fig:exp_decay_Floquet}.
Solid lines show $-E(t)=-\braket{\psi(t)|\hat{H}_0|\psi(t)}$ that is obtained by numerically solving the Schr\"odinger equation starting with the ground state of $\hat{H}_0$ (i.e., the all-down state).
Dashed lines show exponential decays $e^{-\bar{g}t}$ predicted by the Liouvillian gap analysis in the weak dissipation limit.
We find that $\bar{g}$ excellently reproduces the long-time exponential decay in the unitary time evolution.
In this way, thermalization dynamics of an isolated quantum many-body system is better characterized as the weak dissipation limit of the Lindblad dynamics.

\subsection{Dissipator independence of $\bar{g}$}
\label{sec:other}

\begin{figure}
\centering
\includegraphics[width=0.95\linewidth]{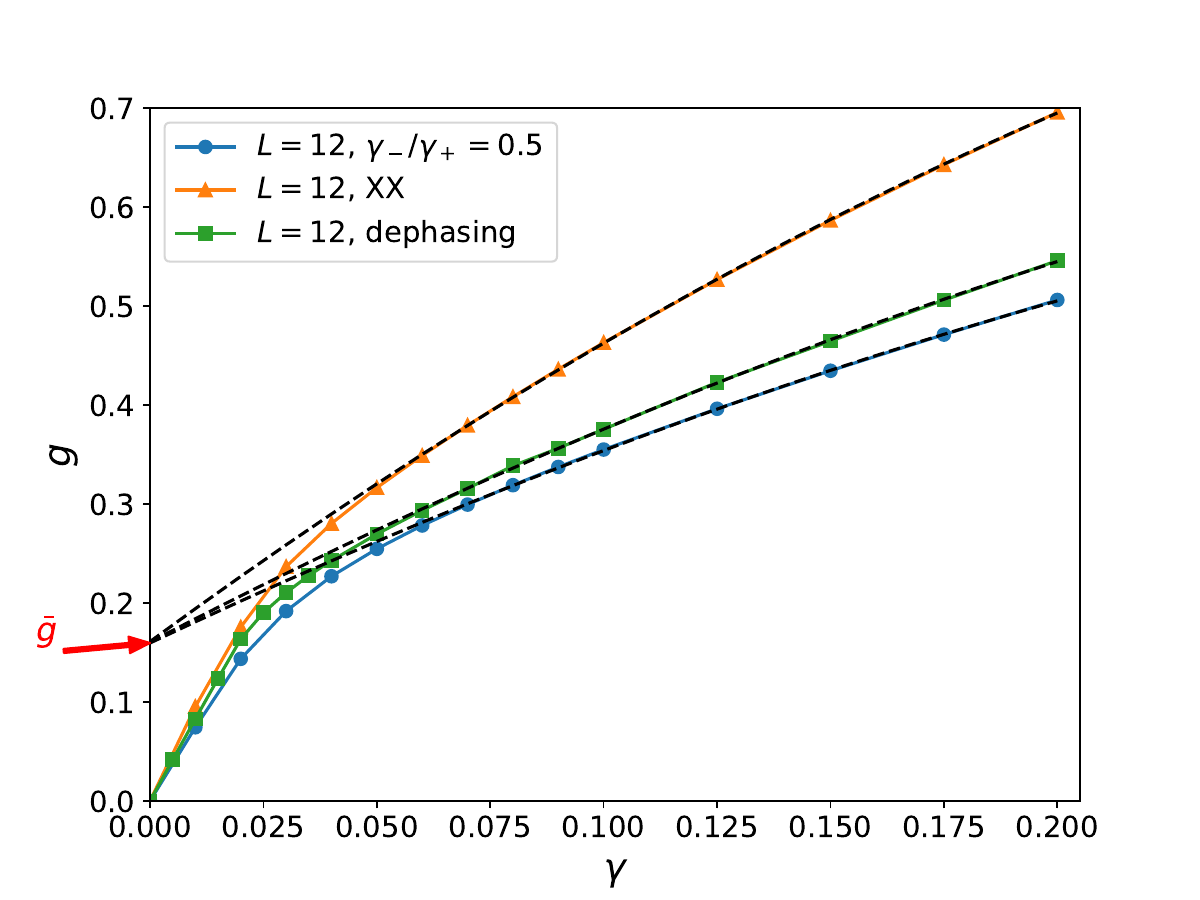}
\caption{Liouvillian gap for the dissipator given by \cref{eq:dissipator_pm} with $\gamma_-/\gamma_+=0.5$ (circles), for the XX dissipator given by \cref{eq:dissipator_XX} (triangles) and for the bulk dephasing (squares). At finite $\gamma$, the Liouvillian gap depends on the dissipator, but the extrapolated value $\bar{g}$ to $\gamma\to +0$ does not.}
\label{fig:updown}
\end{figure}

Up to here, numerical results for the bulk dephasing $\hat{L}_i=\hat{\sigma}_i^z$ have been presented.
Now we numerically show that $\bar{g}$ does not depend on the choice of dissipator.

Let us consider jump operators $\hat{\sigma}_i^+$ and $\hat{\sigma}_i^-$ at every site $i$ with strength $\gamma_+$ and $\gamma_-$, respectively.
The Lindblad equation is given as
\begin{align}
\frac{d\rho}{dt}=&-i[\hat{H}(t),\rho]+\gamma_+\sum_{i=1}^L\left(\hat{\sigma}_i^+\rho\hat{\sigma}_i^--\frac{1}{2}\{\hat{\sigma}_i^-\hat{\sigma}_i^+,\rho\}\right)
\nonumber \\
&+\gamma_-\sum_{i=1}^L\left(\hat{\sigma}_i^-\rho\hat{\sigma}_i^+-\frac{1}{2}\{\hat{\sigma}_i^+\hat{\sigma}_i^-,\rho\}\right),
\label{eq:dissipator_pm}
\end{align}
where $\hat{H}(t)$ is the Hamiltonian of the kicked Ising chain that is given by \cref{eq:H_kick}.
In this paper, we fix the ratio of $\gamma_\pm$ as $\gamma_-/\gamma_+=0.5$ and vary the value of $\gamma\equiv\gamma_+$.

Next, we also consider the ``XX dissipator'' corresponding to the jump operator $\hat{L}_i=\hat{\sigma}_i^x\hat{\sigma}_{i+1}^x$ at each site $i$.
The Lindblad equation reads
\begin{equation}
\frac{d\rho}{dt}=-i[\hat{H}(t),\rho]+\gamma\sum_{i=1}^L\left(\hat{\sigma}_i^x\hat{\sigma}_{i+1}^x\rho\hat{\sigma}_i^x\hat{\sigma}_{i+1}^x-\rho\right).
\label{eq:dissipator_XX}
\end{equation}

In \cref{fig:updown}, we compare the $\gamma$-dependence of the Liouvillian gap for the above two dissipators as well as for the bulk dephasing.
At finite $\gamma$, we see that the Liouvillian gap depends on the dissipator, but the extrapolated value $\bar{g}$ towards $\gamma\to+0$ does not.
This is consistent with the conjecture that $\bar{g}$ describes an intrinsic property of the system, not a property of the dissipator.

\begin{figure*}
\centering
\includegraphics[width=0.85\linewidth]{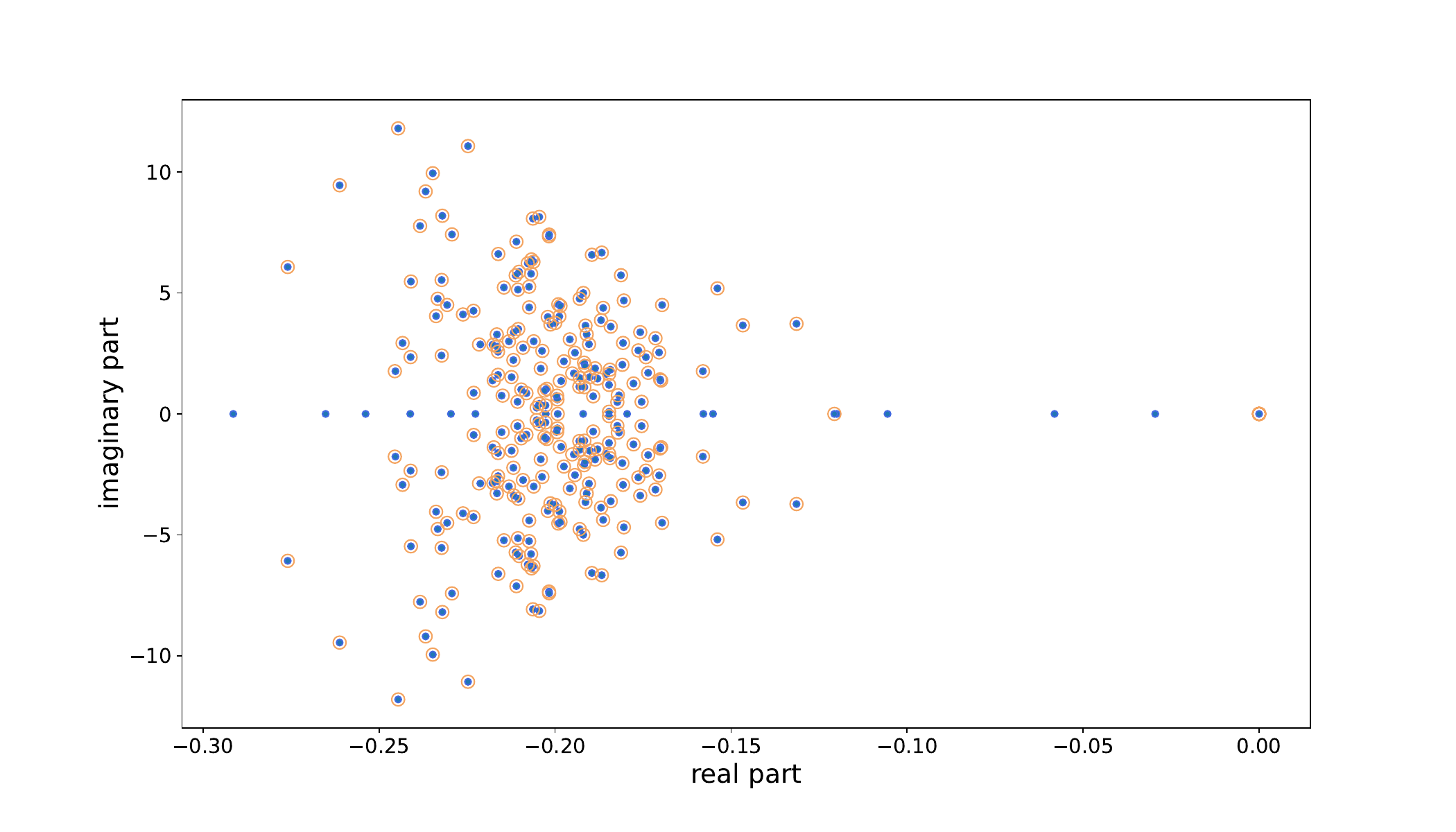}
\caption{Eigenvalues of $\mathcal{L}$ (blue points) and $\mathcal{L}_\mathrm{P}$ (orange circles).}
\label{fig:compare_eigen}
\end{figure*}

\section{Static systems}
\label{sec:static}

In this section, we consider static many-body systems under bulk dissipation.
In contrast to open Floquet systems discussed in \cref{sec:Floquet}, the Liouvillian gap in a static system does not show discontinuity at $\gamma=0$ as is discussed in \cref{sec:no_singularity}.
In \cref{sec:projected}, we explain how to extract the intrinsic decay rate of a static system from the dissipative Liouvillian.
We propose a projection technique which identifies the leading RP resonance in the Liouvillian spectrum: the spectral gap of the projected Liouvillian determines the intrinsic decay rate of a static quantum many-body system.

The discussion below is valid for generic static systems under bulk dissipation, but numerical calculations are carried out for the quantum Ising model under bulk dephasing.
Its Hamiltonian is given by
\begin{align}
\hat{H}=-J\sum_{i=1}^{L-1}\hat{\sigma}_i^z\hat{\sigma}_{i+1}^z-\sum_{i=1}^L\left(h_x\hat{\sigma}_i^x+h_z\hat{\sigma}_i^z\right)
\end{align}
with $(J,h_x,h_z)=(1,0.8090,0.9045)$.
The bulk dephasing is given by $\hat{L}_i=\hat{\sigma}_i^z$ for every site $i$.

\subsection{No singularity in the Liouvillian gap at $\gamma=0$}
\label{sec:no_singularity}

In \cref{sec:Floquet}, it is pointed out that the operator spreading under the unitary time evolution of a Floquet system results in a large asymptotic decay rate under weak bulk dissipation.
More precisely, we have $g\sim L\gamma$ when $\gamma\lesssim v/L$, which implies a singularity of the Liouvillian gap at $\gamma=0$.

However, we now argue that a static system does not exhibit such a singularity.
In \cref{sec:singularity}, we have seen that the operator spreading under the unitary time evolution is a key ingredient to have a large asymptotic decay rate $g\sim L\gamma$ for $\gamma\to +0$ with a fixed system size $L$.
In static systems, the Hamiltonian is invariant under the unitary time evolution at $\gamma=0$, and hence the Hamiltonian does not undergo the operator spreading.
As a consequence, If we consider the effect of weak bulk dissipation, the decay rate of the Hamiltonian under weak bulk dissipation does not increase with time, and the asymptotic decay rate will behave regularly at $\gamma=0$ in the thermodynamic limit.

This argument is consistent with previous studies on the Liouvillian gap in static systems under bulk dissipation~\citep{Shibata2019,Shibata2019_dissipative}.
For example, \citet{Shibata2019} analytically calculated the thermodynamic limit of the Liouvillian gap in the quantum compass model, and obtained $g=2\gamma$ when $\gamma$ is smaller than a certain critical value.
Obviously, $\lim_{\gamma\to+0}\lim_{L\to\infty}g=0$, and thus no singularity appears at $\gamma=0$.
The discontinuity of the Liouvillian gap in the weak dissipation limit is a generic feature of open Floquet systems, but not of open static systems.


In general, when there is a local conserved quantity $\hat{Q}$ in the absence of dissipation (in static systems, we generically have $\hat{Q}=\hat{H}$), the Liouvillian gap does not show singularity in the limit of $\gamma\to +0$ because the conserved quantity does not undergo the operator spreading.
The Liouvillian gap in the weak dissipation regime describes the relaxation of $\hat{Q}$ under dissipation, which is not related to the intrinsic relaxation process of the isolated system.
The Liouvillian gap therefore does not give the leading RP resonance in static systems.

\subsection{Projected Liouvillian and its spectral gap}
\label{sec:projected}

One may ask whether some Liouvillian eigenvalues with smaller real part (i.e. higher decay rates) can be interpreted as RP resonances.
Because the suppression of the singularity is due to the presence of local conserved quantities, it would be expected that one can extract the information of intrinsic decay rates (i.e. RP resonances) by discarding the effect of conserved quantities.
In the energy basis, it is encoded in diagonal matrix elements.
Therefore, we can discard it by applying the following projection superoperator $\mathcal{P}$:
\begin{align}
\mathcal{P}\rho\coloneqq\rho-\sum_n\braket{n|\rho|n}\ket{n}\bra{n},
\end{align}
where $\hat{H}\ket{n}=E_n\ket{n}$.
Let us define the projected Liouvillian as
\begin{align}
\mathcal{L}_\mathrm{P}\coloneqq\mathcal{P}\mathcal{L}\mathcal{P}
\end{align}
and its spectral gap as
\begin{align}
g_\mathrm{P}\coloneqq-\max_{\alpha}\re\lambda_\alpha^\mathrm{P},
\end{align}
where $\lambda_\alpha^\mathrm{P}$ are eigenvalues of $\mathcal{L}_\mathrm{P}$ corresponding to right eigenvectors within the projected subspace, i.e., $\mathcal{L}_\mathrm{P}\rho_\alpha^\mathrm{P}=\lambda_\alpha^\mathrm{P}\rho_\alpha^\mathrm{P}$ and $\mathcal{P}\rho_\alpha^\mathrm{P}=\rho_\alpha^\mathrm{P}$.

In \cref{fig:compare_eigen}, we compare the Liouvillian eigenvalues $\{\lambda_\alpha\}$ and the projected Liouvillian eigenvalues $\{\lambda_\alpha^\mathrm{P}$ in the quantum Ising model under bulk dissipation for $L=4$.
Interestingly, we find that each $\lambda_\alpha^\mathrm{P}$ is almost identical to one of the Liouvillian eigenvalues, and hence we find approximately $\{\lambda_\alpha^\mathrm{P}\}\subset\{\lambda_\alpha\}$.
The projection thus picks up selective eigenvalues of the Liouvillian, and it is plausible to expect that they are related to intrinsic decay rates of the isolated system.
In particular, we expect that the projected Liouvillian gap $g_\mathrm{P}$ corresponds to (real part of) the leading RP resonance.

\subsection{Projected Liouvillian gap as a quantum Ruelle-Pollicott resonance}
\label{sec:static_RP}

\begin{figure}
\centering
\includegraphics[width=0.95\linewidth]{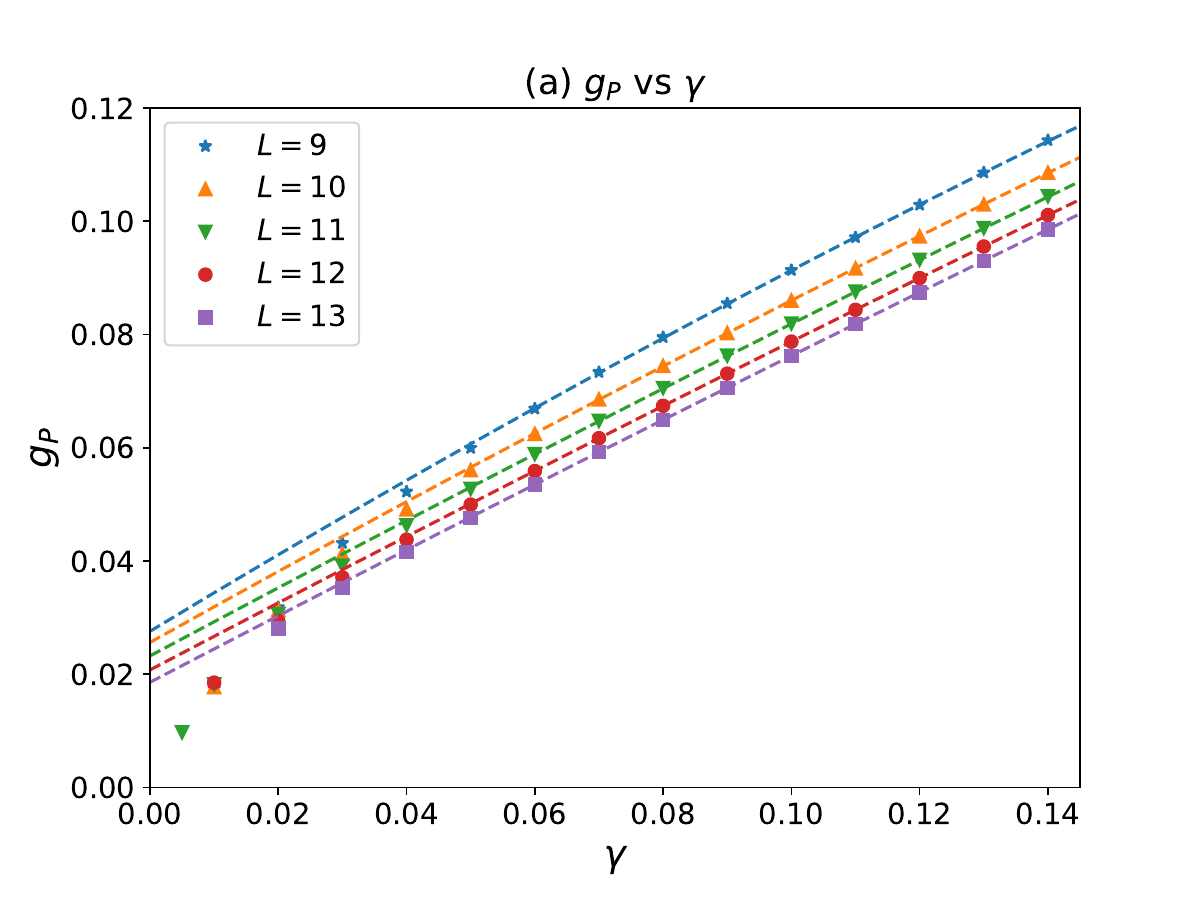}
\includegraphics[width=0.95\linewidth]{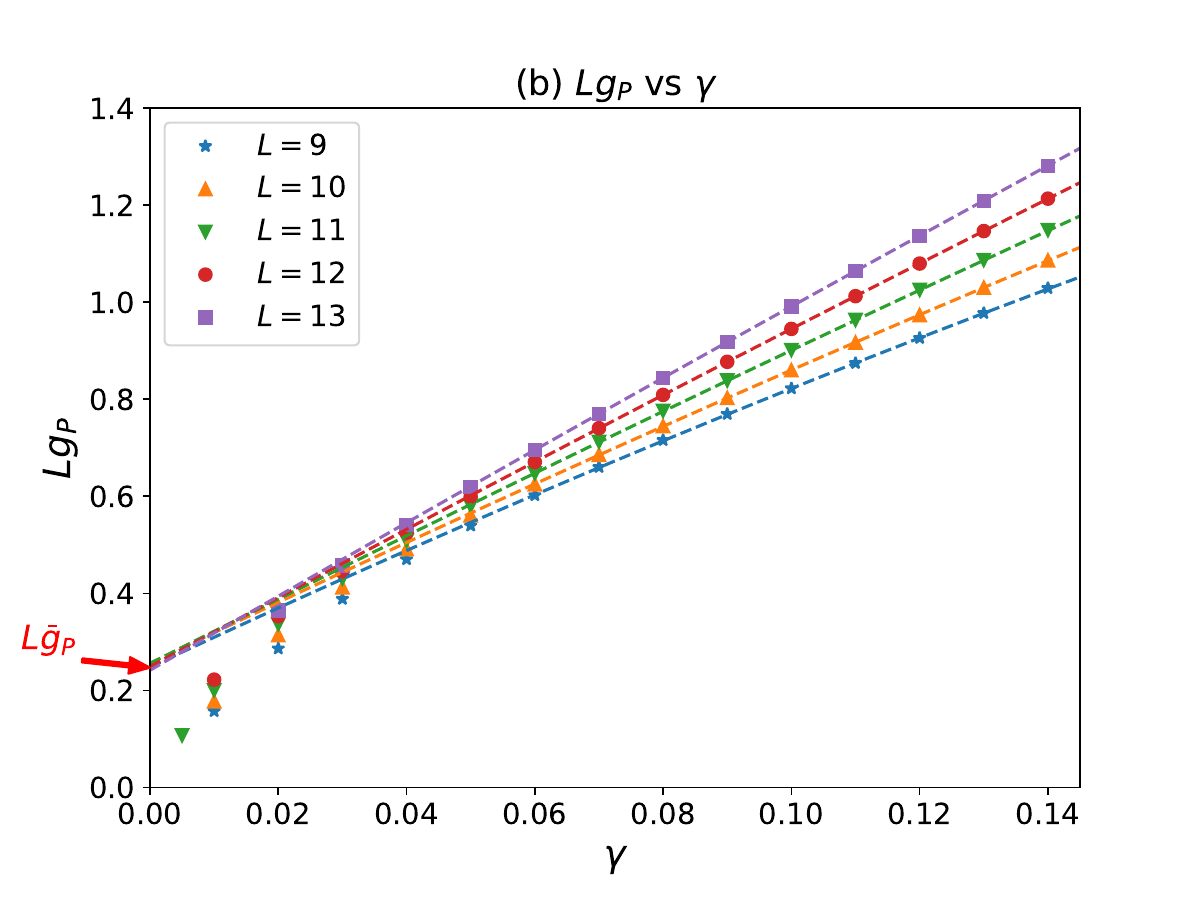}
\caption{Projected Liouvillian gap for various system sizes. (a) $g_\mathrm{P}$ is plotted as a function of $\gamma$. (b) $Lg_\mathrm{P}$ is plotted as a function of $\gamma$. Dashed lines are quadratic functions fitted to the data with $\gamma\geq\gamma^*=0.05$.
In (b), the dashed lines cross at $\gamma=0$, which defines $L\bar{g}_\mathrm{P}$.}
\label{fig:gp}
\end{figure}

We now numerically test our conjecture on the correspondence between the projected Liouvillian gap and the leading RP resonance.
In a static isolated system, the long-time relaxation is governed by hydrodynamic modes which are related to the transport of the energy.
It is expected that the decay rate of the hydrodynamic model of the wave length $L$ vanishes as $L^{-z}$ in the thermodynamic limit, where $z>0$ is the dynamical exponent.
We therefore expect that the leading RP resonance also vanishes in the thermodynamic limit.

We numerically compute the projected Liouvillian gap $g_\mathrm{P}$ as a function of $\gamma$ for various system sizes.
Our numerical results are given in \cref{fig:gp}.
As in Floquet systems, we have a nonzero value by extrapolating the data of $g_\mathrm{P}$ only for $\gamma\geq\gamma_*$ into $\gamma=0$.
This finite value is identified as the leading RP resonance.
In numerical calculations, we set $\gamma_*=0.05$. 

As expected, the extrapolated value $\bar{g}_\mathrm{P}$ decreases as the system size increases.
Our numerical results suggest the scaling $\bar{g}_\mathrm{P}\sim L^{-1}$.
\Cref{fig:gp} (b) demonstrates that the extrapolated curves for different system sizes cross at $\gamma=0$ if we plot $Lg_\mathrm{P}$ as a function of $\gamma$.
This scaling is rather counterintuitive because the diffusive transport of the energy, which occurs in quantum chaotic systems, implies that the decay rate of the slowest hydrodynamic mode is proportional to $L^{-2}$ (i.e. $z=2$).
Indeed, in classical Hamiltonian dynamics, \citet{Gaspard1996} found that in strongly chaotic systems, the leading RP resonance is proportional to $L^{-2}$, which is called the deterministic diffusion.
The scaling $\bar{g}_\mathrm{P}\sim L^{-1}$ indicates the difference between classical and quantum Hamiltonian dynamics.
This apparent discrepancy between transport properties and the dynamical exponent of the projected Liouvillian gap might be related to the gap discrepancy problem in open quantum systems~\citep{Znidaric2015, Mori2020_resolving}.

\begin{figure}
\centering
\includegraphics[width=0.95\linewidth]{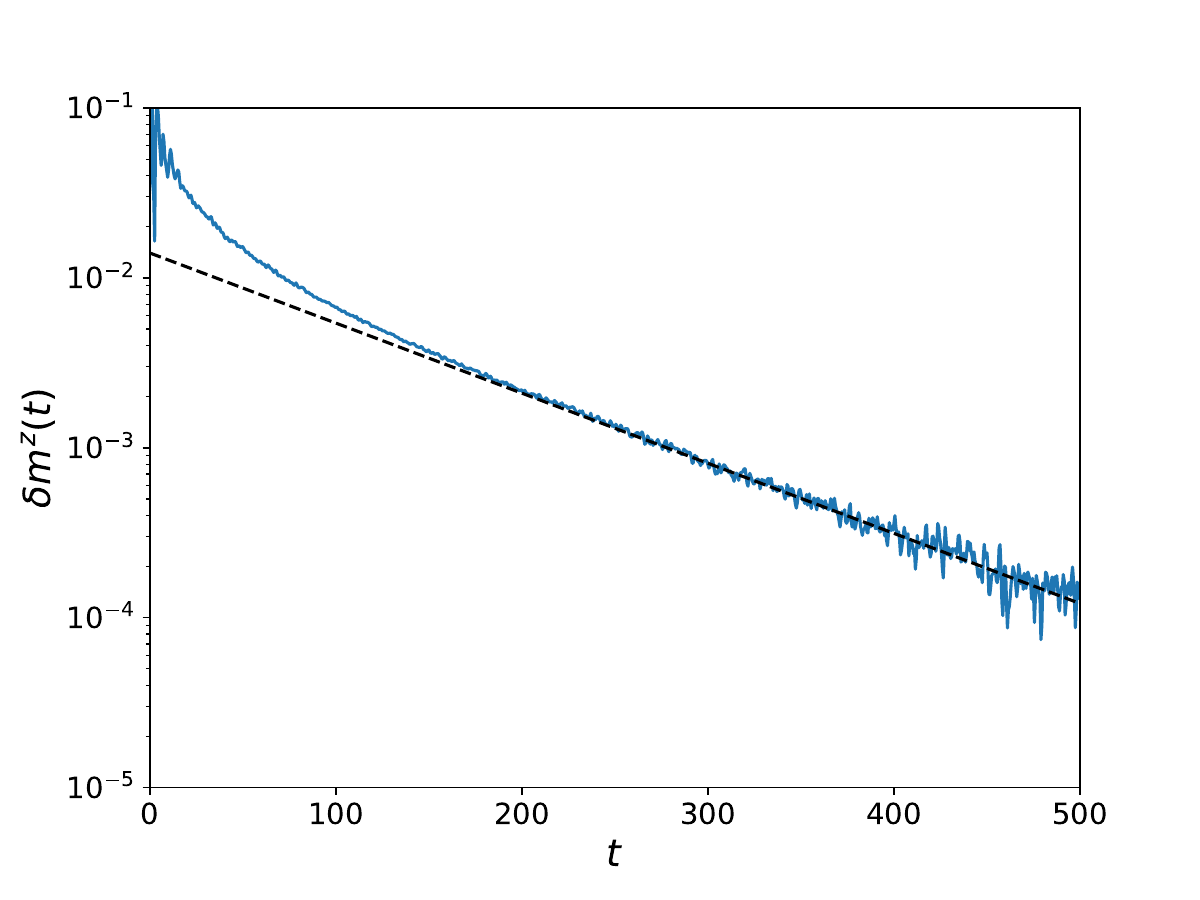}
\caption{Comparison between the numerical solution of the Schr\"odinger equation in the static quantum Ising model with $L=26$ (the solid line) and the exponential decay $\propto e^{-\bar{g}_\mathrm{P}t}$ predicted by the Liouvillian gap analysis in \cref{fig:gp}.}
\label{fig:exp_decay_static}
\end{figure}

The $L^{-1}$ scaling of $\bar{g}_\mathrm{P}$ is more strongly evidenced by comparing the intrinsic decay rate of the isolated system with $\bar{g}_\mathrm{P}$.
In \cref{fig:exp_decay_static}, we plot
\begin{align}
\delta m^z=\frac{1}{L}\left|\braket{\psi(t)|\hat{M}^z|\psi(t)}-\overline{\braket{\psi(t)|\hat{M}^z|\psi(t)}}\right|
\label{eq:delta_mz}
\end{align}
with the solid line, where $\ket{\psi(t)}$ is a numerical solution of the Schr\"odinger equation $id\ket{\psi(t)}/dt=\hat{H}\ket{\psi(t)}$ starting with the all-down initial state for $L=26$, and $\hat{M}^z=\sum_{i=1}^L\hat{\sigma}_i^z$ is the total magnetization.
The overline in \cref{eq:delta_mz} denotes the long-time average.
The dashed line of \cref{fig:exp_decay_static} is $e^{-\bar{g}_\mathrm{P}t}$, where $\bar{g}_\mathrm{P}$ for $L=24$ is estimated as follows: By using numerical data for $L=9$, $10$, $11$, $12$, and $13$, and assuming $\bar{g}_\mathrm{P}\propto L^{-1}$, we obtain $\bar{g}_\mathrm{P}\approx 0.247/L$, which gives an estimate at $L=26$ as $\bar{g}_\mathrm{P}\approx 0.0095$.
We find that the projected Liouvillian gap extrapolated to $\gamma\to +0$ excellently reproduces the intrinsic decay rate.
Thus, in static systems, the projected Liouvillian gap gives the leading RP resonance, and our numerical results strongly support an unexpected scaling $\bar{g}_\mathrm{P}\sim L^{-1}$.


\section{Conclusion}
\label{sec:conclusion}

We have investigated generic properties of the many-body Liouvillian in the weak dissipation regime.
Although recent studies on open quantum many-body systems mainly focus on novel phenomena in the strong dissipation regime, we find that spectral properties of the Liouvillian in the weak dissipation regime has an interesting connection to irreversible relaxation under the intrinsic unitary time evolution of the system.

In Floquet systems under weak bulk dissipation, it turns out that the Liouvillian gap $g$ has discontinuity at $\gamma=0$ in the thermodynamic limit, which is explained by the operator spreading under the intrinsic time evolution.
The nonzero value of $\bar{g}=\lim_{\gamma\to+0}\lim_{L\to\infty}g$ gives the intrinsic decay rate of the isolated system, which is interpreted as the leading Ruelle-Pollicott resonance.
In static systems under weak bulk dissipation, the Liouvillian gap does not show singularity at $\gamma=0$ and is not related to intrinsic irreversible dynamics of the system.
Instead, the projected Liouvillian gap, which is almost identical to another eigenvalue of the original Liouvillian, gives the leading Ruelle-Pollicott resonance.

Those findings clarify unknown general properties of the many-body Lindbladian in the weak dissipation regime, and will trigger further studies on the theory of open quantum many-body systems.
Our work also brings about a new perspective on theoretical description of thermalization of isolated quantum systems.
The Liouvillian-gap analysis discussed in this work allows us to directly access exponentially decaying eigenmodes of an isolated quantum system, which cannot be obtained by just diagonalizing the many-body Hamiltonian.

In this work, we focus on short-range interacting systems.
It is an important future problem to extend the present theory to a wider class of quantum many-body systems.
In particular, long-range interacting systems are paid much attention in recent studies~\citep{Defenu2023,Defenu2023_out}.
Long-range interactions alter the dynamical scaling of the operator spreading~\citep{Yin2020,Guo2020,Kuwahara2021}, which will force us to modify general discussion in \cref{sec:Floquet,sec:static}.

\begin{acknowledgments}
This work was supported by JSPS KAKENHI Grant Numbers JP21H05185 and by JST, PRESTO Grant No. JPMJPR2259.
\end{acknowledgments}

\bibliography{apsrevcontrol, physics}

\end{document}